%% file: main.tex
\documentclass[letterpaper]{article}
\usepackage[preprint]{aaai2027}
\usepackage[hyphens]{url}
\usepackage{graphicx}
\usepackage{natbib}
\usepackage{caption}
\usepackage{algorithm}
\usepackage{algorithmic}
\usepackage{newfloat}
\usepackage{listings}
\DeclareCaptionStyle{ruled}{labelfont=normalfont,labelsep=colon,strut=off}
\floatstyle{ruled}
\newfloat{listing}{tb}{lst}{}
\floatname{listing}{Listing}

\usepackage{amsmath}
\usepackage{amssymb}
\usepackage{booktabs}
\usepackage{xcolor}
\usepackage{tikz}
\usetikzlibrary{positioning,fit,arrows.meta}
\usepackage{fontawesome5}
\usepackage[many]{tcolorbox}

\newcommand{\gainpos}[1]{\textcolor{green!60!black}{+ #1}}
\newcommand{\gainneg}[1]{\textcolor{red}{- #1}}
\input{setup}

\AtBeginDocument{%
  \providecommand\BibTeX{{%
    \normalfont B\kern-0.5em{\scshape i\kern-0.25em b}\kern-0.8em\TeX}}}

\title{TestGRAD: Evolving Test Suites via Failure Pattern Momentum for SWE-Agent Ensemble}

\author{
    Pengfei He\textsuperscript{1},
    Jiayuan Zhou\textsuperscript{2},
    Shaowei Wang\textsuperscript{1}\corresponding,
    Ruiqi Pan\textsuperscript{3}
}
\affiliations{
    \textsuperscript{1}University of Manitoba, Winnipeg, Canada\\
    \textsuperscript{2}Huawei Canada, Markham, Canada\\
    \textsuperscript{3}Huawei Technologies, Hangzhou, China\\
    \{hep2@myumanitoba.ca, shaowei.wang@umanitoba.ca\}\\
    jiayuan.zhou1@huawei.com, panruiqi@huawei.com
}

\begin{document}

\maketitle
\input{0-abstract}

\input{1-introduction}
\input{2-relatedwork}

\input{4-method}

\input{5-experiment}
\input{6-results}

\input{conclusion}

\bibliography{references}

\end{document}


\onecolumn

\maketitle

\small
\raggedbottom
\setlength{\dbltextfloatsep}{12pt plus 2pt minus 2pt}
\setlength{\textfloatsep}{10pt plus 2pt minus 2pt}
\appendix
\input{appendix}

%% file: setup.tex
\usepackage{multirow}
\usepackage{colortbl}
\usepackage{mathtools}
\usepackage{bm}
\usepackage{enumitem}
\usepackage{tabularx}
\usepackage{bbding}
\usepackage{xspace}
\usepackage{pifont}

\newcommand{\lnum}[1]{{\large\textcolor{red!70!black}{\ding{\numexpr191+#1\relax}}}}
\newcommand{\cnum}[1]{{\large\textcolor{teal!80!black}{\ding{\numexpr181+#1\relax}}}}
\newcommand{\ourtool}{\textsc{TestGRAD}\xspace}

%% file: 0-abstract.tex
\begin{abstract}
SWE-agent ensembles improve issue resolution by combining candidate patches from
different agents with complementary strengths. The central problem is therefore
\emph{test-based} selection: generate tests, execute candidate patches, and
identify the best patch. We formulate this process as \emph{test-space
optimization}: evolving an executable repository test suite until it
distinguishes competing patches. Existing test-generation methods are limited
optimizers. They usually lack an explicit loss for ensemble selection, optimize
through incomplete directions that mostly create new tests or delete old ones,
and perform one-off generation without feedback from repeated failures.

Inspired by gradient descent with momentum, we introduce \ourtool, a framework
for automatic test optimization. \ourtool centers on three concepts.
\emph{Differential loss} gives the optimizer an explicit execution-defined
target: useful tests should separate candidate patches by behavior.
\emph{Full CRUD gradients} expand the update direction from merely creating or
deleting tests to reading existing test infrastructure, creating new tests,
updating stale assertions, and deleting only obsolete tests. \emph{Failure
Pattern Momentum} mines frequent failure sequences from memory, allowing the
optimizer to avoid repeated non-discriminative directions while compressing the
failure-history context. On SWE-bench Verified, \ourtool achieves
\textbf{84.2\%} Pass@1 with a 4-agent ensemble, outperforming the strongest
baseline (80.6\%) by an absolute improvement of \textbf{3.6 percentage points},
while compressing failure-history context by over \(100{\times}\).
\end{abstract}

%% file: 1-introduction.tex
\section{Introduction}
\label{sec:introduction}
Software engineering (SWE) agents autonomously resolve repository issues by
localizing faults, generating patches, and validating fixes in a live coding
environment~\citep{SWEbench,Lingxi}. Different agents exhibit complementary
strengths: an agent that fails on one issue may still produce the best patch for
another. A practical way to exploit this diversity is parallel ensembling, where
multiple SWE agents generate candidate patches and a selector chooses the final
patch~\citep{tts}.

The selector must be grounded in behavior. A natural but weak approach is to ask
an LLM judge to compare candidate patches and select the most plausible
fix~\citep{Augment,Deibase}. Without execution evidence, however, the judge sees
surface plausibility rather than repository behavior and can be sensitive to
presentation order~\citep{bias1,biasacl}. The key to SWE-agent ensembling is
therefore \emph{test-based selection}: generate tests, execute every candidate
patch, and use the resulting behavior to identify the correct patch.

This view turns ensemble selection into 
\emph{test-space optimization}: the optimized object is the repository test
suite, and the objective is to evolve it for better selection. This is difficult for three reasons. First, the loss is
not obvious. A test that simply passes or fails is not necessarily useful for
selection; even patches that pass benchmark tests can remain behaviorally
divergent from developer fixes~\citep{patchdiff}. Second, the update direction
cannot be limited to creating new tests. Useful repository tests often require
reading fixtures and conventions, updating stale assertions, and deleting only
confirmed obsolete tests~\citep{swtbench}. Third, test generation is rarely solved in one step:
when an attempted test is non-discriminative, the next attempt should use that
failure as optimization signal rather than restart from scratch.

Inspired by gradient descent with momentum in neural networks~\citep{polyak1964momentum,sutskever2013momentum}, we
introduce \ourtool, a framework that models test-suite evolution as
gradient-descent-like optimization in test space. Table~\ref{tab:analogy}
summarizes the analogy. A test suite is the parameter state, differential
execution is the loss, CRUD edits are gradient directions, frequent failure
sequences are momentum, support is the retention weight, and the LLM agent is
the optimizer. In the forward pass, \ourtool executes the current test suite on
candidate patches and observes whether their behaviors differ. In the backward
pass, the LLM optimizer uses this execution-defined signal and mined failure
momentum to produce the next CRUD test update. The design has three components.

\lnum{1} \textbf{Differential loss.} \ourtool makes the optimization target
explicit: a useful test suite should separate candidate patches by behavior. We
define the loss by execution. If the current test suite contains a
differential test, meaning at least one test passes on some candidate patches and
fails on others, the loss reaches zero. Otherwise, the loss remains nonzero and
the failed update is used to guide the next iteration. This differs from
pass-count selection~\citep{Agentless,trae}, which simply chooses the patch that
passes the most tests. Differential loss instead searches for behavioral
contrast among candidates rather than merely verifying the intended fix. It
prunes candidates that are indistinguishable under the current tests and
discourages selection based on a single weak pass/fail signal.

\lnum{2} \textbf{Full CRUD gradient.} \ourtool treats test-suite evolution as a
full CRUD update rather than a Create/Delete-only pipeline. Existing
test-generation methods~\citep{Agentless,trae} mainly create a new test from the
issue and, at most, delete conflicting tests by name matching. This partial
direction is insufficient for repository tests, which often depend on existing
fixtures, setup routines, imports, assertion helpers, and tests whose expected
values must be revised after the fix.

\ourtool therefore exposes all four directions as the test-space gradient.
\textbf{\underline{R}ead} extracts repository-specific testing knowledge for
reuse. \textbf{\underline{C}reate} adds new discriminative tests.
\textbf{\underline{U}pdate} revises stale assertions or test bodies whose
expected behavior changes with the fix. \textbf{\underline{D}elete} removes only
tests confirmed obsolete. This full direction avoids shallow standalone tests,
preserves high-signal tests that only need assertion updates, and prevents a
correct patch from being rejected simply because an old assertion was left
unchanged.

\lnum{3} \textbf{Failure Pattern Momentum.} In numerical optimization, momentum
accumulates consistent gradient information across steps and helps avoid
oscillation or repeated dead ends. \ourtool uses an analogous mechanism for test
updates. When a CRUD update produces a non-discriminative suite, the failed
update is written into failure memory. \ourtool then mines maximal frequent
failure sequences from this memory. These sequences form Failure Pattern
Momentum: compact evidence of directions that have repeatedly failed. The next
LLM optimizer step receives this momentum and can move away from recurring
non-discriminative regions of the test-suite space. Compared with summarizing
failures using another LLM call~\citep{reflexion} or feeding raw test-execution
feedback back to the generator~\citep{agentcoder}, this design is deterministic,
faithful to observed failures, and compresses context by over $100{\times}$.

On SWE-bench Verified (500 instances)~\citep{swebenchleaderboard}, \ourtool
achieves \textbf{84.2\%} Pass@1 with a 4-agent ensemble using Minimax-m2.7,
outperforming the next-best baseline (TRAE, 80.6\%) by an absolute improvement
of \textbf{3.6 percentage points} and the single-agent baseline by
\textbf{5.0 percentage points}. Gains are consistent across all three base LLMs.
Ablations confirm the contribution of each component: removing Read
drops Pass@1 by $3.0\%$, removing Update by $2.4\%$, replacing differential
loss with pass-count selection by $3.4\%$, removing Failure Pattern Momentum by
$1.6\%$, and removing iterative optimization entirely by $9.2\%$.

\noindent\textbf{Contributions:}
\begin{enumerate}[leftmargin=2em,topsep=2pt,itemsep=2pt,label=\cnum{\arabic*}]
  \item We formulate SWE-agent ensemble selection as test-space optimization over executable repository test suites.
  \item We instantiate this optimization with differential execution loss and full CRUD test-suite gradients.
  \item We introduce Failure Pattern Momentum, which mines frequent failure sequences from memory to guide iterative test evolution while compressing context by over $100{\times}$.
  \item \ourtool achieves 84.2\% Pass@1 on SWE-bench Verified with a 4-agent ensemble.
\end{enumerate}

%% file: 2-relatedwork.tex
\section{Related Work}
\label{sec:related}
\input{Table/baselines}
\subsection{Test Generation for SWE-Agent Ensemble}

LLM-based agents have transformed automated program repair, with recent systems
resolving over half of real-world GitHub issues on SWE-bench~\citep{SWEbench}.
Representative frameworks include Agentless~\citep{Agentless}, which decomposes
repair into localization, patch generation, and validation, OpenHands~\citep{OpenHands},
which enables interactive repository navigation with tool use, and Lingxi~\citep{Lingxi},
which guides issue resolution using procedural knowledge from past fixes.
Because different SWE agents exhibit complementary strengths~\citep{Deibase},
ensembling their candidate patches is a natural way to improve Pass@1. The key
question is how to generate tests that can distinguish the correct patch from
plausible but wrong alternatives.

Table~\ref{tab:analogy} summarizes the analogy that motivates \ourtool:
tests are optimized as text-space parameters, CRUD edits instantiate the gradient
direction, frequent failure sequences provide momentum, and the LLM agent acts as
the optimizer.

\paragraph{Create/Delete test pipelines.}
LIBRO~\citep{libro} generates a reproduction test directly from the issue
description and ranks candidate patches by pass/fail outcome. Agentless~\citep{Agentless}
follows a related \emph{Create-Delete} pipeline: it creates a new test from the
issue description and removes conflicting tests by name matching. These pipelines
provide useful execution evidence, but they are incomplete as test optimizers in
three ways. First, their update direction is mostly limited to creating tests,
with deletion used only as a coarse cleanup step. They do not read and update the
existing test suite, so they miss project-specific fixtures, base classes, helper
assertions, and stale assertions that need in-place revision. Second, their
objective is not a differential loss over candidate patches. A test may pass or
fail, but the generator is not directly optimized to maximize behavioral
separation among competing patches. Third, they are largely one-off: once the
first generated test is non-discriminative, there is no principled iterative
optimization process that turns that failure into the next test-suite update.

\paragraph{Agentic test generation.}
AEGIS~\citep{AEGIS} uses FSM-guided script generation to reproduce bug symptoms
as standalone scripts. These scripts verify whether the symptom disappears, but
they do not integrate with the project's native test suite and cannot fully
capture behavioral regression. TRAE~\citep{trae} adds agentic browsing over
source files to inform test creation, but it still follows a Create/Delete style
of test generation, does not optimize a differential loss, and does not maintain
momentum across failed test updates. \ourtool extends this line of work by
treating the test suite itself as the optimized object: it reads existing tests,
updates stale assertions, creates new discriminative tests, deletes only confirmed
obsolete tests, and iteratively evolves the suite using Failure Pattern Momentum.

\subsection{Reflection and Text-Space Optimization}

Reflexion~\citep{reflexion} and Self-Refine~\citep{selfrefine} introduced the paradigm of linguistic self-reflection: an agent receives execution feedback, generates a natural-language reflection, and uses it to guide subsequent attempts.
Language Agent Tree Search~(LATS)~\citep{mtcs} scales this idea with Monte Carlo Tree Search, using LM-powered value functions to explore the space of possible actions.
These frameworks rely on an LLM evaluator to produce the reflection signal, which is costly and may hallucinate for fine-grained code-correctness reasoning.
AgentCoder~\citep{agentcoder} instead uses a test executor agent that runs generated tests and feeds the raw execution results back to the programmer agent for refinement, without abstracting recurring failures across attempts.

TextGrad~\citep{textgrad} gives this idea a more explicit optimization form:
LLM feedback is backpropagated through text variables in a computation graph, so
prompts, predictions, code snippets, or other textual components can be improved
by natural-language gradients. metaTextGrad~\citep{metatextGrad} further studies
how to optimize the LLM optimizer itself, adapting both its prompt and structure
for a target task. These works motivate our use of gradient-style language for
test-suite optimization, but they are general text-optimization frameworks rather
than a test-time SWE-agent ensemble method. In our setting, the forward signal is
produced by executing candidate patches against tests, the loss is differential
over patch outcomes, and the gradient direction is constrained to CRUD edits over
the repository test suite.

\ourtool takes a different approach to memory utilization.
Rather than summarizing failed attempts with an LLM~\citep{reflexion} or feeding raw execution feedback back verbatim~\citep{agentcoder}, it applies sequential mining to failure memory.
This produces compact maximal frequent subsequences that represent structurally identical strategies across superficially diverse patches.
In \ourtool, these sequences serve as Failure Pattern Momentum: they carry aggregate information about prior non-discriminative directions into the next LLM optimizer step.
Unlike similarity-based retrieval~(RAG), which returns documents similar to the query and therefore also similar to each other, sequential mining extracts cross-document structure and compresses memory by over $100{\times}$ in practice, keeping the optimizer's context window available for productive repository reasoning.


%% file: Table/baselines.tex
\begin{table}[tbp]
\centering
\scriptsize
\caption{Role-level analogy between gradient descent with momentum and
test-space optimization. A one-step CRUD edit plays the role of an
instantaneous gradient, whereas a high-support recurring failed edit sequence plays
the role of momentum carried across attempts.}
\label{tab:analogy}
\setlength{\tabcolsep}{4pt}
\renewcommand{\arraystretch}{1.15}

\begin{tabularx}{\columnwidth}{
  @{}>{\raggedright\arraybackslash}X
     >{\raggedright\arraybackslash}X@{}}
\toprule
\textbf{Gradient Descent with Momentum}
&
\textbf{\ourtool}
\\
\midrule

\textcolor{gray!70!black}{\faCode}~
\textbf{Parameter state.}
Current parameter vector \(\theta_t\).
&
\textcolor{gray!70!black}{\faCode}~
\textbf{Test suite state.}
Current candidate test suite \(y_t\).
\\[2pt]

\textcolor{gray!70!black}{\faCompass}~
\textbf{Objective.} 
Loss function \(\mathcal{L}(\theta_t)\).
&
\textcolor{gray!70!black}{\faCompass}~
\textbf{Objective.} 
Differential loss \(\mathcal{L}(y_t)\) measuring whether the current test suite has produced a differential outcome.
\\[2pt]

\textcolor{gray!70!black}{\faFilter}~
\textbf{Retention strength.}
\(\beta \in[0,1]\) controls how much of the previous velocity is retained.
&
\textcolor{gray!70!black}{\faFilter}~
\textbf{Support-based retention.}

The \(\displaystyle \beta=\operatorname{MinSupport}\) threshold defines the minimum frequency a sequence must satisfy. Lowering this threshold retains more sequences and gives momentum broader influence.
\\[2pt]

\textcolor{gray!70!black}{\faHistory}~
\textbf{Momentum.}
\(\displaystyle \beta \nabla_{t-1}\).
The accumulated direction from previous steps.
&
\textcolor{gray!70!black}{\faHistory}~
\textbf{Momentum analogue.}
\(\displaystyle m_t=\operatorname{Mine}_{\beta}(M)\).
A persistent test update distilled from recurring failure sequences in memory \(M\), after the latest failed update has been appended.
\\[2pt]

\rowcolor{yellow!10}
{\faChartLine}~
\textbf{Update with Momentum.}
\(\displaystyle \nabla_t=\beta \nabla_{t-1}+(1-\beta)\frac{\partial \mathcal{L}}{\partial \theta_t}\).
Combines instantaneous gradient with momentum to suppress noisy one-step gradients and reinforce consistent directions.
&
{\faChartLine}~{\textbf{Gradient analogue.}}
\(\displaystyle \nabla_{\text{TestGrad}}=\operatorname{Optimizer}(\mathcal{L},y_t,m_t)\).
Combines current feedback with momentum to suppress one-off noisy edits and reinforce consistent test-improvement directions.
\\[2pt]

\textcolor{gray!70!black}{\faEdit}~
\textbf{State update.}
\(\displaystyle \theta_{t+1}=\theta_t-\eta \nabla_t\).
&
\textcolor{gray!70!black}{\faEdit}~
\textbf{Test update.}
\(\displaystyle y_{t+1}=y_t+\nabla_{\text{TestGrad}}\),
where \(\nabla_{\text{TestGrad}}\) is a combination of CRUD operations on the original test suite.
\\

\bottomrule
\end{tabularx}
\end{table}

%% file: 4-method.tex
\section{TestGRAD Framework}
\label{sec:method}

\begin{figure}[tbp]
\centering
\includegraphics[width=\columnwidth]{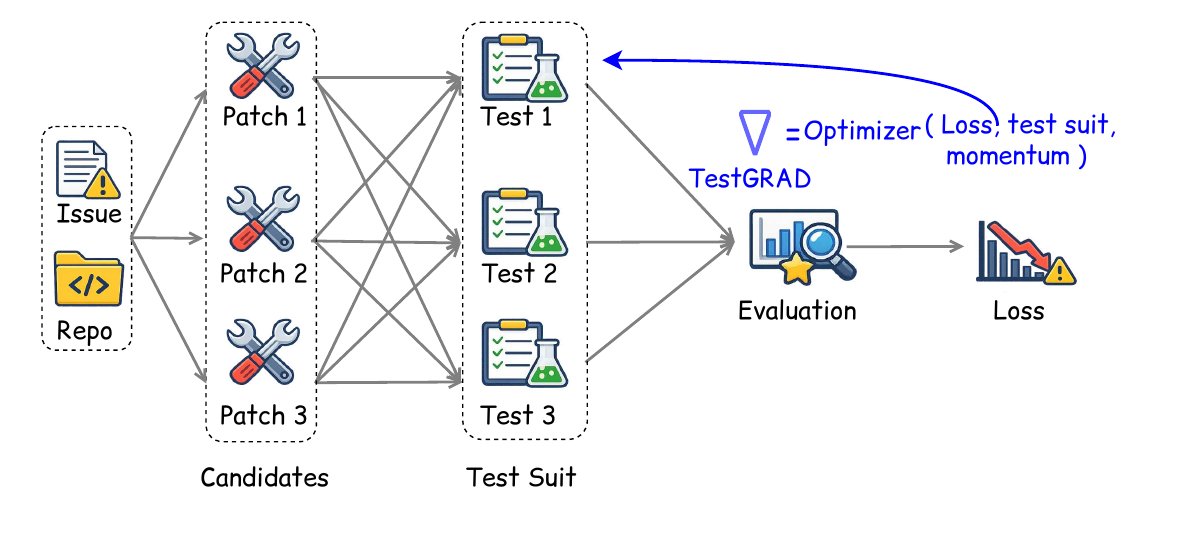}
\caption{\ourtool framework overview. Given an issue and candidate patches from
multiple SWE agents, \ourtool iteratively evolves the repository test suite:
the forward pass executes tests to obtain differential loss; failed
non-differential updates are stored in failure memory and mined into
Failure Pattern Momentum; the backward pass uses the loss and momentum to
produce the next CRUD test update.}
\label{fig:testgrad-framework}
\end{figure}

Table~\ref{tab:analogy} gives the role-level analogy that structures
\ourtool. A test suite is the optimization state, differential execution is the
loss, a one-step CRUD edit is the instantaneous gradient, and frequent failure
sequences mined from memory are the momentum carried across attempts. The whole
method follows the same loop: execute the current test suite, compute the
differential loss, update failure memory and momentum, and then back-propagate
the signal into the next CRUD test update. Figure~\ref{fig:testgrad-framework}
shows this end-to-end framework, Figure~\ref{fig:testgrad-gradient} highlights
the derivative step implemented by the LLM optimizer, and
Figure~\ref{fig:workflow-blocks} zooms into the generic transition from
\(y_t\) to \(y_{t+1}\) through failure memory and momentum.

\subsection{Problem and Test-Space State}
\label{sec:formulation}

Let \(c=(I,\mathcal{R}_0)\) denote a SWE-bench instance, where \(I\) is the
issue description and \(\mathcal{R}_0\) is the initial repository. A SWE-agent
ensemble \(\{\pi_1,\ldots,\pi_N\}\) produces candidate patches
\(\mathcal{P}=\{p_i\}_{i=1}^N\). Applying patch \(p_i\) to \(\mathcal{R}_0\)
yields a candidate repository. The optimization state is the current test suite
\(y_t\). The initial state \(y_0\) is the original repository test suite before
applying the issue-specific update.

\subsection{Forward Pass: Execute Candidate Patches}
\label{sec:forward}

The forward pass is purely execution-based. Given the current test suite
\(y_t\), \ourtool runs it against every candidate patch in \(\mathcal{P}\) and
records the pass/fail outcomes:
\begin{equation}
    y_t
    \xrightarrow{\;\textsc{Eval}(\cdot,\mathcal{P})\;}
    \text{pass/fail outcomes}.
    \label{eq:forwardgraph}
\end{equation}
A test suite is useful for ensemble selection only if at least one test is
\emph{differential}: it passes on some candidate patches and fails on others.
Uniform outcomes do not separate candidates and therefore provide no selection
signal.

\subsection{Loss: Differential Execution}
\label{sec:loss}

The loss measures whether the current test suite has produced a differential
outcome:
\begin{equation}
    \mathcal{L}(y_t)=
    \begin{cases}
    0, & \text{if } y_t \text{ contains a differential test},\\
    1, & \text{otherwise}.
    \end{cases}
    \label{eq:loss}
\end{equation}
This is an execution-defined loss, following the differential-testing intuition
that behavioral differences across executions expose semantic
distinctions~\citep{differential}. When \(\mathcal{L}(y_t)=0\), optimization
stops immediately. The checker agent is then invoked once, outside the loop and
with a separate context, to use the differential result for final patch
selection. When \(\mathcal{L}(y_t)=1\), the failed update becomes the
optimization signal for the next iteration.

\subsection{Memory Update and Momentum}
\label{sec:mining}

If the current test suite is non-differential, the CRUD update that produced it
has failed to separate candidate patches. \ourtool writes this failed update
into failure memory \(M\):
\begin{equation}
    M \leftarrow M \mathbin{\|} [\nabla_{t-1}].
    \label{eq:memory}
\end{equation}
Here \(\nabla_{t-1}\) is the one-step CRUD update that produced the current failed
suite \(y_t\), and \(\|\) appends it to memory so repeated failures remain available for
support counting. Instead of replaying all raw failures to the next iteration,
\ourtool mines maximal frequent failure sequences from memory:
\begin{equation}
    m_t=\operatorname{Mine}_{\beta}(M).
    \label{eq:momentum}
\end{equation}
The support threshold \(\beta\) (default \(\beta=2\)) plays the role of retention strength in
Table~\ref{tab:analogy}. Lowering \(\beta\) retains more recurring failure
sequences, giving momentum broader influence; raising \(\beta\) keeps only the
most stable recurring sequences. Thus \(m_t\) is a compact, deterministic momentum
signal: it suppresses one-off failed edits and records which directions have
already been repeatedly explored, so the next update can move away from the
same dead end.

\subsection{Backward Pass: CRUD Test Update}
\label{sec:crud}

The backward pass sends the loss and momentum back to the LLM optimizer. The
optimizer produces the next test-space gradient:
\begin{equation}
    \nabla_{\text{TestGrad}}^t
    =
    \operatorname{Optimizer}(\mathcal{L},y_t,m_t).
    \label{eq:testgradgrad}
\end{equation}
Figure~\ref{fig:testgrad-gradient} expands this step as a test-space derivative:
the optimizer prompt receives the current suite, the execution-defined loss, and
the mined momentum, then returns a CRUD edit that plays the role of
\(\partial\mathcal{L}/\partial y_t\).
This gradient is a CRUD edit over the original repository test suite. Read
collects repository-specific testing knowledge such as fixtures, imports,
assertion utilities, and setup conventions. Create adds new tests, Update
revises stale assertions or test bodies, and Delete removes tests confirmed
obsolete. Following the textual-gradient view~\citep{textgrad,metatextGrad},
these CRUD edits are the gradient direction in text/code space rather than a
numerical derivative.

The state update is:
\begin{equation}
    y_{t+1}
    =
    y_t+\nabla_{\text{TestGrad}}^t .
    \label{eq:update}
\end{equation}
The new suite \(y_{t+1}\) enters the next forward pass. The loop continues until
a differential test is found or the iteration budget is exhausted.
Figure~\ref{fig:workflow-blocks} illustrates this transition: repeated
non-discriminative failures update memory and momentum, and the optimizer
converts \((\mathcal{L},y_t,m_t)\) into a CRUD edit that yields \(y_{t+1}\).

\input{Figure/testgrad_gradient_box}
\input{framework}

%% file: Figure/testgrad_gradient_box.tex
\begin{figure*}[tbp]
\centering
\begin{tcolorbox}[
  enhanced,
  width=0.98\textwidth,
  colback=blue!2,
  colframe=blue!65!black,
  colbacktitle=blue!65!black,
  coltitle=white,
  boxrule=0.9pt,
  arc=2pt,
  left=8pt,
  right=8pt,
  top=6pt,
  bottom=6pt,
  title={\large\textbf{Gradient computation for test-space optimization:}
  \(y_t \xrightarrow{\;\textsc{Eval}\;} \mathcal{L}(y_t)
  \xrightarrow{\;\textsc{Optimizer}\;} \nabla_{\text{TestGrad}}^t\)}
]

\small
\begin{center}
\begin{minipage}{0.92\textwidth}
\[
  \frac{\partial \mathcal{L}}{\partial y_t}
  \;\equiv\;
  \nabla_{\textsc{LLM}}
  \bigl(y_t,\mathcal{L}(y_t),m_t\bigr)
  \;\approx\;
  \nabla_{\text{TestGrad}}^t
\]
\vspace{-0.5em}
\[
  \nabla_{\text{TestGrad}}^t
  =
  \operatorname{Optimizer}\bigl(\mathcal{L},y_t,m_t\bigr),
  \qquad
  y_{t+1}=y_t+\nabla_{\text{TestGrad}}^t .
\]
\end{minipage}
\end{center}

\vspace{0.2em}
\noindent\textbf{Prompt template implementing the derivative.}
\begin{tcolorbox}[
  enhanced,
  colback=white,
  colframe=gray!55,
  boxrule=0.45pt,
  arc=1pt,
  left=6pt,
  right=6pt,
  top=4pt,
  bottom=4pt
]
\footnotesize\ttfamily
Given issue \{I\}, current test suite \{y\_t\}, execution result
\{loss L(y\_t)\}, and momentum \{m\_t\}:\\
1. Explain why the current suite is still non-differential.\\
2. Identify the failed direction preserved in momentum.\\
3. Return one CRUD update over the original repository tests:\\
\hspace*{1em}Read required test infrastructure; Create new discriminative tests;\\
\hspace*{1em}Update stale assertions or test bodies; Delete only obsolete tests.
\end{tcolorbox}

\vspace{0.2em}
The operator \(\nabla_{\textsc{LLM}}\) returns natural-language feedback and a
concrete CRUD edit rather than a numeric derivative. The differential loss
provides the downstream objective, and \(m_t\) carries recurring failed
directions from memory, so the optimizer can update the test suite away from
previous non-discriminative attempts.
\end{tcolorbox}
\caption{\ourtool gradient computation in the backward pass. The derivative
step is implemented by an LLM optimizer that converts execution-defined loss
and Failure Pattern Momentum into a CRUD update over the repository test suite.}
\label{fig:testgrad-gradient}
\end{figure*}

%% file: framework.tex
%

\definecolor{cissue}{HTML}{FFF3CD}
\definecolor{cpatch}{HTML}{D6E4FF}
\definecolor{cresult}{HTML}{D1FAE5}
\definecolor{cfailed}{HTML}{FEE2E2}
\definecolor{cmomentum}{HTML}{EDE9FE}

\tcbset{
  wfblock/.style={
    enhanced,
    colframe=black!80,
    boxrule=0.9pt,
    arc=2.5pt,
    top=1mm, bottom=1mm, left=1.5mm, right=1.5mm,
    before upper={\setlength{\parskip}{0pt}\setlength{\partopsep}{0pt}},
  },
  wfissue/.style   = {wfblock, colback=cissue},
  wfpatch/.style   = {wfblock, colback=cpatch},
  wfresult/.style  = {wfblock, colback=cresult},
  wffailed/.style  = {wfblock, colback=cfailed},
  wfmomentum/.style = {wfblock, colback=cmomentum},
}

\lstset{basicstyle=\fontsize{5.35}{6.2}\selectfont\ttfamily, breaklines=true, frame=none, numbers=none, xleftmargin=0pt, xrightmargin=0pt, escapechar=~, escapebegin=\normalfont, showstringspaces=false}

\newcommand{\wfcap}[2]{%
  \vspace{1pt}%
  \centering{\scriptsize\rmfamily\textbf{#1}~#2}%
}

\begin{figure}[tbp]
\centering
\newlength{\wfblkw}
\setlength{\wfblkw}{0.96\columnwidth}
\begingroup
\lstset{basicstyle=\fontsize{5.35}{6.2}\selectfont\ttfamily, breaklines=true, frame=none, numbers=none, xleftmargin=0pt, xrightmargin=0pt, escapechar=~, escapebegin=\normalfont, showstringspaces=false}

\begin{minipage}{\wfblkw}
  \begin{tcolorbox}[wfpatch]
    \begin{lstlisting}[language=Python]
# ~\faSearch~ Read and Reuse: app/warning fixtures
# ~\faWrench~ current suite y_t
def test_toctree_special_pages(
    app, status, warning):
  app.build()
  warnings = warning.getvalue()
  assert "...genindex'" not in warnings
  assert "...modindex'" not in warnings
  assert "...search'" not in warnings
  assert 'index' in app.env.toctree_includes
    \end{lstlisting}
  \end{tcolorbox}
  \wfcap{(a)}{Current suite $y_t$: warning check}
\end{minipage}

\vspace{1mm}

\begin{minipage}{\wfblkw}
  \begin{tcolorbox}[wfmomentum]
    \begin{lstlisting}
~\faChartBar~ Momentum(M): |M|=3 failed suites
-> ~{\setlength{\fboxsep}{0.6pt}\colorbox{yellow!55}{\strut\textbf{retained frequent sequence}}}~
-> all converge on warning checking
~\textcolor{orange!75!black}{\textbf{[frequent: support 3/3 $\geq \beta{=}2$]}}~
~{\setlength{\fboxsep}{0.4pt}\colorbox{yellow!55}{\strut\ttfamily app.build()}}~
~{\setlength{\fboxsep}{0.4pt}\colorbox{yellow!55}{\strut\ttfamily warn = warning.getvalue()}}~
~{\setlength{\fboxsep}{0.4pt}\colorbox{yellow!55}{\strut\ttfamily assert "...genindex'" not in warn}}~
~{\setlength{\fboxsep}{0.4pt}\colorbox{yellow!55}{\strut\ttfamily assert "...modindex'" not in warn}}~
~{\setlength{\fboxsep}{0.4pt}\colorbox{yellow!55}{\strut\ttfamily assert "...search'"   not in warn}}~

~\textcolor{red!70!black}{\faSkull}~ ~\textbf{Dead end:}~ "assert no warning" exhausted (3/3 attempts)
~\textcolor{green!60!black}{\faLightbulb}~ ~\textbf{Steer:}~ inspect AST via assert_node()
    \end{lstlisting}
  \end{tcolorbox}
  \wfcap{(b)}{Momentum $m_t=\textsc{Mine}(M)$}
\end{minipage}

\vspace{1mm}

\begin{minipage}{\wfblkw}
  \begin{tcolorbox}[wfpatch]
    \begin{lstlisting}[language=Python]
# ~\faLightbulb~ ~$y_{t+1} = y_t + \nabla_{\text{TestGrad}}^t$~
# ~\faArrowRight~ avoid dead-end warning-check
# ~\faStar~ new: probe toctree AST structure
def test_toctree_index(app):
    app.build()
    toctree = app.env.tocs['index']
    node = toctree[0][1][1]
    assert_node(node, addnodes.toctree,
        caption="Indices",
        entries=[(None, 'genindex'),
                 (None, 'modindex'),
                 (None, 'search')])
    \end{lstlisting}
  \end{tcolorbox}
  \wfcap{(c)}{Updated suite $y_{t+1}$: AST probe}
\end{minipage}

\caption{%
  \ourtool momentum-gradient transition from $y_t$ to $y_{t+1}$, illustrated on
  \texttt{sphinx-doc\_\_sphinx-10673}.
  \textbf{(a)}~The current suite $y_t$ checks warning text and yields uniform outcomes
  ($\mathcal{L}(y_t)=1$).
  \textbf{(b)}~Failure memory is mined into $m_t$; the highlighted statements form the frequent
  sequence shared by all three failed suites (support $3\geq\beta$), marking this repeated failed
  direction as a dead end and steering the optimizer toward AST inspection.
  \textbf{(c)}~The updated suite $y_{t+1}$ probes toctree structure with
  \texttt{assert\_node()}, producing a discriminative signal for the checker.
  Colors:
  \colorbox{cpatch}{\strut~test~}
  \colorbox{cmomentum}{\strut~momentum~}.%
}
\label{fig:workflow-blocks}
\endgroup
\end{figure}

%% file: 5-experiment.tex
\section{Experimental Setup}
\label{sec:setup}

\paragraph{Metrics.} \textbf{Pass@1}: Fraction of the 500 instances resolved under the SWE-bench evaluation harness, after \ourtool's checker selects one patch per instance. \textbf{Ensemble Gain}: Absolute improvement in Pass@1 over the best single-agent patch source in our candidate pool (Livesweagent, 79.2\%).

\paragraph{Dataset.}
We evaluate on \textbf{SWE-bench Verified}~\citep{swe.bench.verified}, a human-validated subset of 500 real-world GitHub issues.
Candidate patches are drawn from six top-performing SWE-agent leaderboard submissions, added incrementally to form ensembles of size $N=1,\ldots,6$:
Livesweagent~(20251215)~\citep{livesweagent}, Sonar~(20251205)~\citep{sonar}, OpenHands~(20251127)~\citep{OpenHands},
Trae\_doubao~(20250928)~\citep{trae}, Gemini-Live~(20251120)~\citep{livesweagent}, and Atlassian~(20250902)~\citep{Atlassian}.
At $N=1$ (Livesweagent alone), the single-agent Pass@1 is 79.2\%, which serves as the baseline for ensemble gain.
This construction has two advantages:
(1)~leaderboard agents produce high-quality and diverse patches, making ensemble selection challenging;
(2)~all patches are publicly available, ensuring full reproducibility.

\paragraph{Baselines.}
We compare \ourtool against three ensemble methods. For a fair comparison, all methods receive the same candidate patches and differ only in their ensemble selection module.
\textbf{Augment}~\citep{Augment} follows the \emph{LLM-as-a-judge} paradigm: the LLM scores each candidate patch against the issue description and selects the most plausible fix without any test execution. \textbf{Agentless}~\citep{Agentless} generates a new test from the issue description and filters existing tests by name-matching heuristics, then ranks patches by execution outcome. It follows a \emph{Create-Delete~(C/D)} strategy with no repository inspection and no iterative optimization. \textbf{TRAE}~\citep{trae} is an agent-based selector that adds repository browsing to the C/D pipeline.
It reads source files to inform test creation and applies name-based deletion to remove conflicting existing tests, but still lacks an Update operation and does not refine failing tests.

\paragraph{Implementation.}
\ourtool uses three base LLMs interchangeably as its LLM optimizer: \textbf{Minimax-m2.7}~\citep{minimax}, \textbf{DeepSeek-v3.2}~\citep{deepseek}, and \textbf{GLM-5.1}~\citep{glm5}.
The maximum number of optimization steps is $T_{\max}=8$.
Failure Pattern Momentum uses sequential mining with a minimum support threshold of $\beta=2$ (a sequence must appear in at least 2 failed test suites to be reported).
All runs execute inside the official SWE-bench Docker images.



%% file: 6-results.tex
\section{Results}\label{sec:results}
\input{rq1}
\input{rq2}
\input{rq3}
\input{rq4}

%% file: rq1.tex
\subsection{Overall Effectiveness of \ourtool}
\label{sec:rq1}

Table~\ref{tab:main} summarizes the Pass@1 performance of all ensemble methods across three base LLMs on SWE-bench Verified.
\ourtool consistently achieves the highest Pass@1 in every configuration:
84.2\% with Minimax-m2.7 (\gainpos{5.0}),
84.0\% with DeepSeek-v3.2 (\gainpos{4.8}),
and 83.6\% with GLM-5.1 (\gainpos{4.4}).

\textbf{Augment} fails to improve over single-agent performance and even degrades it
(\gainneg{0.6} to \gainpos{0.0}), confirming that LLM-as-a-judge scoring is
unreliable for patch selection due to positional bias and the absence of execution
grounding. \textbf{Agentless} and \textbf{TRAE} show modest improvements of up to
\gainpos{1.2} and \gainpos{1.4}, respectively.
Both rely on a Create-Delete (C/D) pipeline: they generate new tests while
filtering existing ones by test name, lacking holistic test suite
understanding (Read) and in-place test updating (Update).
\ourtool surpasses both by 3.4--3.6\%, demonstrating the
clear benefit of evolving test suites with CRUD gradients and momentum-guided
optimization.

\input{Table/main_table}

%% file: Table/main_table.tex
\begin{table*}[!htbp]
\centering
\scriptsize
\caption{Performance comparison. Gain is computed relative to the baseline Pass@1 (79.2\%). Positive gains are shown in green, while negative gains are shown in red.}
\begin{tabular}{lcccccccc}
\toprule
\multirow{2}{*}{\textbf{Base LLM}} & \multicolumn{2}{c}{\textbf{Augment}} & \multicolumn{2}{c}{\textbf{Agentless}} & \multicolumn{2}{c}{\textbf{TRAE}} & \multicolumn{2}{c}{\textbf{Ours}} \\
\cmidrule(lr){2-3} \cmidrule(lr){4-5} \cmidrule(lr){6-7} \cmidrule(lr){8-9}
& Pass@1 (\%) & Gain (\%) & Pass@1 (\%) & Gain (\%) & Pass@1 (\%) & Gain (\%) & Pass@1 (\%) & Gain (\%) \\
\midrule
Minimax-m2.7  & 79.0 & \gainneg{0.2} & 80.4 & \gainpos{1.2} & 80.6 & \gainpos{1.4} & 84.2& \gainpos{5.0}  \\
GLM-5.1       & 78.6 & \gainneg{0.6} & 80.2 & \gainpos{1.0} & 80.2 & \gainpos{1.0} & 83.6&  \gainpos{4.4}  \\
DeepSeek-v3.2 & 79.2 & \gainpos{0.0} & 79.8 & \gainpos{0.6} & 80.6 & \gainpos{1.4} & 84.0&  \gainpos{4.8}  \\
\bottomrule
\end{tabular}
\label{tab:main}
\end{table*}

%% file: rq2.tex
\subsection{Impact of Ensemble Size}
\label{sec:rq2}

\begin{figure}[tbp]
\centering
\includegraphics[width=\linewidth]{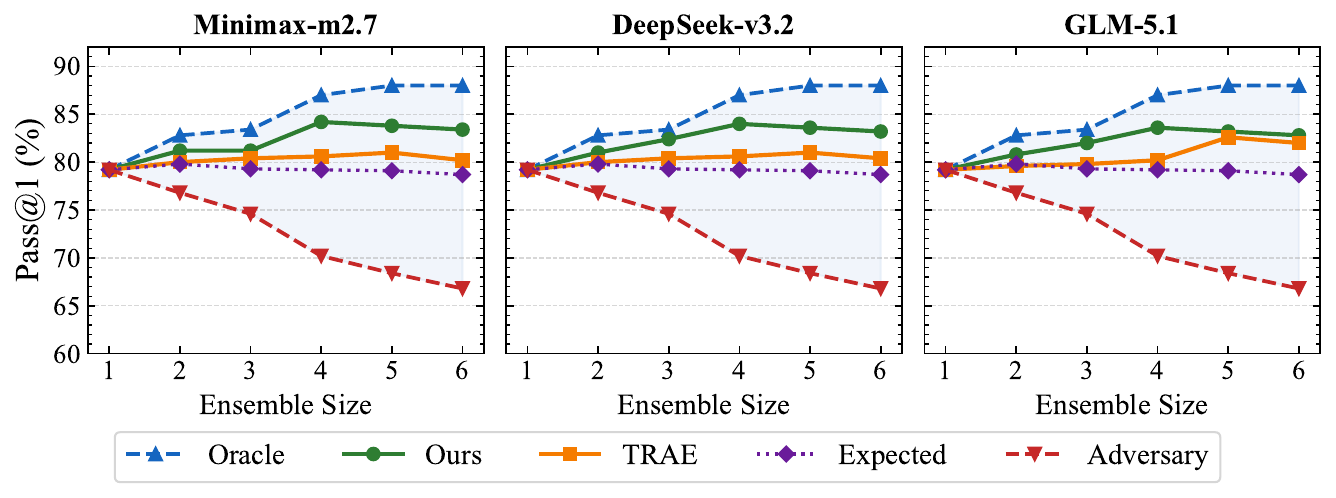}
\caption{Pass@1 vs.\ ensemble size for \ourtool (green) and TRAE
(orange), together with Oracle (blue dashed), Adversary (red dashed), and
Expected (purple dotted) bounds. Results shown for Minimax, DeepSeek, and GLM
(left to right).}
\label{fig:ensemble_size}
\end{figure}

Figure~\ref{fig:ensemble_size} plots Pass@1 as a function of ensemble size $N$
for \ourtool and TRAE across three LLMs, together with the Oracle and Adversary bounds and the Expected
(random-selection) baseline.
Table~\ref{tab:ensemble_minimax} reports the full numerical results for Minimax-m2.7.

\ourtool consistently outperforms TRAE at every ensemble size $N{\geq}2$.
With Minimax the gap widens from $+1.2\%$ at $N{=}2$ to $+3.6\%$ at $N{=}4$,
showing that CRUD-gradient updates and momentum become increasingly valuable
as the selection task grows harder with more diverse candidates.

\textbf{Optimal ensemble size.}
Performance peaks at $N{=}4$ (Minimax: 84.2\%, DeepSeek: 84.0\%, GLM: 83.6\%)
and slightly declines for $N{=}5,6$.
This trend should not be interpreted as a violation of the ensemble principle.
When the underlying selection loss is non-convex, increasing ensemble size can
produce plateaus, elbows, or even local non-monotonicity rather than a strictly
increasing performance curve~\citep{jmlr}.
The decline is consistent with the Adversary and Expected curves, which also
decrease beyond $N{=}4$ (Expected: 79.2\% $\to$ 78.7\%): adding weaker
agents (Gemini-Live at 77.4\%, Atlassian at 76.8\%) increases candidate noise
and lowers the probability that a simple selector chooses a correct patch, even
though the Oracle ceiling can still rise.
Despite this, \ourtool remains well above the Expected baseline for all $N{\geq}2$,
confirming it extracts reliable discriminative signal even from large candidate sets.

At $N{=}4$ the Oracle is 87.0\%, leaving a 2.8\% gap to \ourtool (84.2\%).
This gap represents instances where a correct patch exists but the generated
tests fail to discriminate it from incorrect candidates.

\begin{table}[tbp]
\centering
\scriptsize
\setlength{\tabcolsep}{3pt}
\caption{Pass@1 across ensemble sizes (\ourtool, Minimax-m2.7).}
\begin{tabular}{lccccc}
\toprule
\textbf{Agent} & \textbf{N} & \textbf{Adv.} & \textbf{Exp.} & \textbf{Oracle} & \textbf{P@1} \\
\midrule
Livesweagent  & 1 & 79.2 & 79.2 & 79.2 & 79.2 \\
+Sonar        & 2 & 76.8 & 79.8 & 82.8 & 81.2 \\
+OpenHands    & 3 & 74.6 & 79.3 & 83.4 & 81.2 \\
+Trae\_doubao & 4 & 70.2 & 79.2 & 87.0 & \textbf{84.2} \\
+Gemini       & 5 & 68.4 & 79.1 & 88.0 & 83.8 \\
+Atlassian    & 6 & 66.8 & 78.7 & 88.0 & 83.4 \\
\bottomrule
\end{tabular}
\label{tab:ensemble_minimax}
\end{table}

\begin{table}[tbp]
\centering
\scriptsize
\setlength{\tabcolsep}{4pt}
\caption{Ablation study on SWE-bench Verified (Minimax-m2.7).}
\begin{tabular}{@{}lp{1.3in}|cc@{}}
\toprule
\textbf{Model} & \textbf{Setting} & \textbf{P@1} & \textbf{$\Delta$} \\
\midrule
\ourtool & Complete & \textbf{84.2} & \textcolor{gray}{--} \\
\midrule
$\ourtool_{\text{w/o Read}}$   & w/o suite understanding & 81.2 & \textcolor{red}{-3.0} \\
$\ourtool_{\text{w/o Update}}$ & w/o updating tests      & 81.8 & \textcolor{red}{-2.4} \\
$\ourtool_{\text{w/o R+U}}$    & w/o Read and Update     & 79.4 & \textcolor{red}{-4.8} \\
\midrule
$\ourtool_{\text{w/o Diff Loss}}$ & pass-count selection   & 80.8 & \textcolor{red}{-3.4} \\
$\ourtool_{\text{w/o Momentum}}$ & w/o failure momentum   & 82.6 & \textcolor{red}{-1.6} \\
$\ourtool_{\text{w/o Opt}}$      & w/o iterative optimizer & 75.0 & \textcolor{red}{-9.2} \\
\bottomrule
\end{tabular}
\label{tab:ablation}
\end{table}


%% file: rq3.tex
\subsection{Contribution of Differential Loss and CRUD Gradients}
\label{sec:rq3}

Table~\ref{tab:ablation} reports the ablation results on Minimax-m2.7.
We focus here on differential loss and the two CRUD-gradient components unique
to \ourtool: \emph{Read} (test suite understanding) and \emph{Update}
(in-place test modification).
Because the complete system and each ablation are evaluated on the same
SWE-bench Verified instances, we use a paired hypothesis test rather than
compare unpaired aggregate Pass@1 values. Specifically, we collect multiple
paired overall Pass@1 results for the complete \ourtool and each ablation
variant, and apply a two-sided Wilcoxon signed-rank test~\citep{wilcoxon} at
significance level $\alpha=0.05$. The complete \ourtool shows a statistically
significant advantage over every ablated variant, with all $p$-values below
$7.32\times10^{-3}$.
The Pass@1 drops in Table~\ref{tab:ablation} are then interpreted as effect
sizes.
For the non-differential-loss ablation, we replace the differential stopping
criterion and checker-based final selection with a pass-count rule: after
executing the generated tests on all candidate patches, the candidate that
passes the largest number of tests wins. This variant tests whether ordinary
test-pass maximization is sufficient, as opposed to explicitly optimizing for a
test whose outcomes differ across candidates. Recent empirical evidence on
SWE-bench further cautions against this assumption: patches that pass benchmark
tests can still be behaviorally divergent from developer-written fixes, and
differential patch testing exposes discrepancies missed by ordinary validation
tests~\citep{patchdiff}.

\textbf{Effect of Differential Loss.}
Replacing differential loss with pass-count selection reduces Pass@1 from
84.2\% to 80.8\% ($\Delta=-3.4\%$). This shows that simply rewarding the patch
that passes the most generated tests is not enough for ensemble selection:
pass count measures compatibility with the current generated tests, not
agreement with the intended fix. Because tests are not exhaustive, a broadly
plausible but incorrect patch may pass many weak tests, while a correct patch can
fail a stale or mis-specified one. The differential objective instead searches
for tests that separate candidate behavior before the checker is invoked.

\textbf{Effect of Read.}
Removing test suite understanding ($\ourtool_{\text{w/o Read}}$) drops Pass@1
from 84.2\% to 81.2\% ($\Delta = -3.0\%$), the largest CRUD-component drop.
Without Read, the agent cannot discover existing fixtures, base test classes, or
helper functions.
Generated tests therefore rely on bare Python imports and ad-hoc setups that
frequently fail to run in the project's test runner or exercise shallower
code paths than tests built on top of the project's own infrastructure.
This confirms that \emph{understanding the repository's testing conventions is
the most critical prerequisite for a useful test-suite gradient}.

\textbf{Effect of Update.}
Removing in-place test updating ($\ourtool_{\text{w/o Update}}$) drops Pass@1
from 84.2\% to 81.8\% ($\Delta=-2.4\%$). When Update is disabled, tests that
assert \emph{old} expected values are deleted rather than corrected. This breaks
regression checks, since these tests still encode the correct behavioral
constraints and only require updated assertions. The $-2.4\%$ drop shows that
deletion-only handling introduces wrong selection and weakens the differential
signal of the test suite.

\textbf{Joint effect of Read and Update.}
Removing both components ($\ourtool_{\text{w/o R+U}}$) further reduces Pass@1
to 79.4\% ($\Delta=-4.8\%$). This variant is close to a pure
Create/Delete-style test generator: it neither grounds new tests in the
repository's existing test infrastructure nor repairs stale assertions in
existing tests. The larger combined drop shows that Read and Update are
complementary CRUD directions. Read supplies the project-specific context needed
to make generated tests executable, while Update preserves and corrects
high-signal tests that would otherwise be discarded.

%% file: rq4.tex
\subsection{Impact of Failure Pattern Momentum and Token Efficiency}
\label{sec:rq4}

We evaluate the momentum mechanism from two angles:
(1) its contribution to Pass@1 via the ablation study, and
(2) its token efficiency advantage over raw memory accumulation.

\textbf{Performance impact.}
From Table~\ref{tab:ablation},
$\ourtool_{\text{w/o Momentum}}$ drops Pass@1 from 84.2\% to 82.6\%
($\Delta = -1.6\%$).
While the gap is smaller than Read or Update in isolation, Failure Pattern Momentum provides
a qualitatively different benefit: it operates \emph{across} iterations, distilling
accumulated failure history into compact momentum that steers the optimizer away from
dead ends.
The larger baseline gap of $-9.2\%$ for $\ourtool_{\text{w/o Opt}}$ (75.0\%)
confirms that iterative optimization as a whole is essential, and momentum is the
mechanism that makes this optimization efficient rather than redundant.

\begin{figure}[!htbp]
\centering
\includegraphics[width=\linewidth]{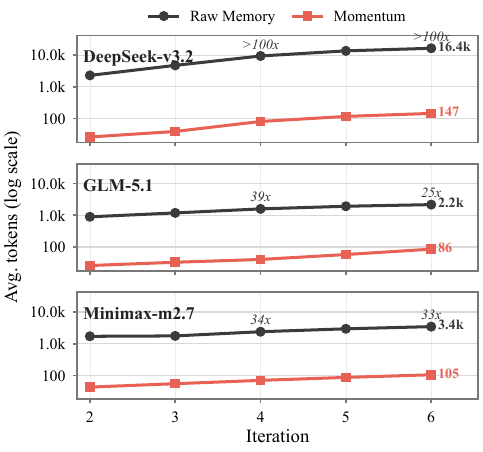}
\caption{Average token consumption of Raw Memory (dark gray) vs.\ Momentum
(coral) across optimization iterations, shown on a log scale for DeepSeek,
GLM, and Minimax. Italic labels denote representative compression ratios
(Raw / Momentum); ratios above \(100{\times}\) are capped as
\emph{\({>}100x\)}.}
\label{fig:compression}
\end{figure}

\textbf{Token efficiency.}
Figure~\ref{fig:compression} illustrates the token cost of the two strategies across optimization iterations.
\emph{Raw Memory} accumulates all previously failed test suites verbatim;
its token count grows steadily, reaching about 1.9K--13.6K tokens by iteration~5.
\emph{Momentum} extracts maximal frequent subsequences from the same
corpus, producing a concise momentum context that grows far more slowly, remaining
on the order of tens to about a hundred tokens across iterations.
The compression ratio exceeds $100{\times}$ for DeepSeek at later iterations
and stays between $25{\times}$ and $40{\times}$ for GLM and Minimax, meaning the model receives equivalent
or richer structural guidance at a fraction of the context cost.
This is particularly important because excessive context length degrades LLM
reasoning quality~\citep{lost} and inflates API costs.

%% file: conclusion.tex
\section{Conclusion}
\label{sec:conclusion}

We presented \ourtool, a momentum-gradient optimizer for iterative test-space
optimization. Following the analogy in Table~\ref{tab:analogy}, \ourtool treats
the current test suite as the optimization state, differential execution as the
loss, a one-step CRUD test update as the instantaneous gradient, and
Failure Pattern Momentum as the retained direction. The LLM optimizer uses
this feedback to evolve discriminative test suites, while the checker is
invoked only after zero-loss differential execution for final patch selection.

On SWE-bench Verified, \ourtool achieves 84.2\% Pass@1 with a 4-agent ensemble,
outperforming the strongest baseline by 3.6\%. Failure Pattern Momentum also
acts as a compact retention mechanism, preserving repeated failed update
directions from failure memory while reducing raw failure-history context by over
$100{\times}$.

%% file: appendix.tex

\input{3-motivation}

\section{Human--LLM Collaboration for Aspect Discovery in Test Suite Understanding}
\label{app:aspects}

\paragraph{What the Read step collects.}
In the Method section, the Read step collects repository-specific test
infrastructure before any CRUD update is written. In practice, this means that
the optimizer first extracts a compact checklist of reusable testing knowledge
from the target repository. The checklist has six categories:
\begin{itemize}[leftmargin=1.5em,topsep=2pt,itemsep=1pt]
  \item project API imports and module paths;
  \item project-specific assertion utilities and call signatures;
  \item \texttt{setUp}/\texttt{setUpTestData} bodies and \texttt{conftest.py} fixtures;
  \item test base classes and inheritance structure;
  \item expected literal values and behavioral contracts from similar tests;
  \item shared helper or factory functions defined in the test module.
\end{itemize}
These aspects are derived empirically from SWE-bench training instances, as
described below.

\paragraph{Methodology.}
We sample 200 instances from SWE-bench, excluding SWE-bench Verified to avoid
data leakage. For each instance, Minimax-m2.7 analyzes the ground-truth test file
and summarizes the test-suite understanding aspects.

Aspects present in fewer than 50\% of instances are discarded.
This threshold ensures that retained aspects are sufficiently \emph{pervasive}
across repositories; aspects below this threshold tend to reflect
repository-specific idioms (e.g., custom database fixtures in a single project)
and do not generalize well.
The remaining six aspects form the final taxonomy used by the Read step.

\paragraph{Human--LLM validation.}
To assess annotation reliability, one of the authors independently re-labels a
random subset of 40 instances (20\% of the sample) without access to the LLM
outputs.
Agreement between the human annotator and the LLM reaches
a Cohen's $\kappa$ score of 0.82, indicating almost perfect agreement and
suggesting that the LLM annotations do not introduce systematic bias.

\paragraph{Findings.}
Table~\ref{tab:aspects} reports the frequency of each retained aspect.
All six aspects appear in over 60\% of test files, confirming that they capture
pervasive conventions rather than repository-specific idioms.
\emph{Project APIs / Imports} and \emph{Assertion Conventions} are the most universal
($>90\%$), as nearly all test files import project modules and rely on
project-specific assertion utilities (e.g., \texttt{assertQuerysetEqual},
\texttt{assert\_array\_equal}).
\emph{Fixtures} and \emph{Test Base Classes} appear in roughly 80\% of instances,
reflecting the prevalence of Django-style \texttt{TestCase} inheritance and pytest
\texttt{conftest.py} conventions.
\emph{Expected Values} and \emph{Helper Functions} are present in about two-thirds
of test files, typically in suites with shared data builders or custom matchers.

\begin{table}[H]
\centering
\footnotesize
\caption{Frequency of the six retained test-suite understanding aspects across 200
sampled SWE-bench instances (SWE-bench Verified excluded).
Frequency is the fraction of instances where the aspect is present in the target
test file. The six discarded aspects each fell below 50\%.}
\setlength{\tabcolsep}{5pt}
\begin{tabular}{llcc}
\toprule
\textbf{Aspect} & \textbf{Typical evidence} & \textbf{Freq.\ (\%)} & \textbf{Cohen's $\kappa$} \\
\midrule
Project APIs / Imports
  & \texttt{from django.core.checks import Error}
  & 96 & 0.91 \\
Assertion Conventions
  & \texttt{self.assertQuerysetEqual(...)}, \texttt{assert\_array\_equal}
  & 91 & 0.88 \\
setUp / Fixtures
  & \texttt{setUp()}, \texttt{setUpTestData()}, \texttt{conftest.py}
  & 83 & 0.84 \\
Test Base Classes
  & \texttt{class MyTest(SimpleTestCase):}
  & 79 & 0.86 \\
Expected Values
  & literals in \texttt{assertEqual}, docstring examples
  & 71 & 0.79 \\
Helper Functions
  & \texttt{make\_response()}, \texttt{assert\_called\_with(...)}
  & 64 & 0.76 \\
\bottomrule
\end{tabular}
\label{tab:aspects}
\end{table}

These six aspects instantiate the \textbf{Read} operation used by \ourtool.
By requiring the agent to locate repository infrastructure before any CRUD
update is written, \ourtool grounds generation in the project's own abstractions
rather than reconstructing them from scratch, the failure mode of C/D pipelines
discussed in the Method section.

\section{Overhead Analysis}
\label{app:overhead}
\input{Table/effciency}
Table~\ref{tab:cost} reports average per-instance time and token consumption
across all four methods and three base LLMs.

\paragraph{Sources of overhead.}
\ourtool's token cost is higher than all baselines for two structural reasons.
First, the Read step issues repository exploration calls once per generation
round to extract test-suite understanding.
Second, \ourtool may run up to $T_{\max}=8$ momentum-guided optimization steps
per instance, which raises the average token cost to 670K--1{,}670K per instance.
By contrast, Augment and Agentless are one-shot LLM calls, and TRAE runs a
single agentic pass without iterative feedback, so their token budgets are
structurally lower.
This overhead is the direct cost of two design choices: holistic test-suite
understanding and iterative momentum optimization, which account for most of the
accuracy gain.

\paragraph{Time.}
Despite the higher token count, \ourtool's time is shorter than TRAE on all
three base LLMs (e.g., 12.2 vs.\ 19.3 min on Minimax-m2.7), because optimization steps run sequentially
only when needed and early stopping applies once a discriminative test is found.
Agentless and Augment are faster but operate in a one-shot regime, foregoing the
$+9.2\%$ gain that iterative optimization provides.

\paragraph{Accuracy--cost tradeoff.}
For applications where selection quality is critical, where a wrong patch
propagates a bug to production, the additional token cost translates directly
into +5.0\% Pass@1 over the single-agent baseline and +3.6\% over the next-best
method. Reducing overhead further (e.g., by capping $T_{\max}$ or pruning the
Read step) is straightforward at the cost of proportionally lower accuracy, and
we leave adaptive budget allocation to future work.

%% file: 3-motivation.tex
\section{Motivation}
\label{sec:motivation}
In this section, we illustrate why CRUD-gradient updates over the original test
suite are necessary, and why Failure Pattern Momentum is essential for effective
iterative optimization.

\input{Figure/motivation_graph}

\subsection{CRUD Gradient over Test Suites}

Figure~\ref{fig:motivation} contrasts two test generation strategies on
\texttt{django\_\_django-12325}, a bug where \texttt{OneToOneField} to a
parent model should implicitly set \texttt{parent\_link=True} instead of
raising \texttt{ImproperlyConfigured}.
The fix changes an exception-raising behavior into a silently-accepted one,
making test design non-trivial: a test must assert the \emph{new} correct
behavior, not merely reproduce the old crash.
We use this instance to identify two critical gaps in existing C/D-only
(Create + Delete) approaches.

\label{sec:motivation-crud}
\paragraph{Read.}
Generating a test without first inspecting the repository forces the agent
to reconstruct from scratch everything the project already provides.
As shown in Figure~\ref{fig:motivation}(A), the C/D-only reproduction script
must manually configure \texttt{DJANGO\_SETTINGS\_MODULE}, call
\texttt{django.setup()}, and declare \texttt{app\_label} in every model's
\texttt{Meta} class, boilerplate that Django's own test suite has long
abstracted away behind the \texttt{@isolate\_apps} decorator.
Beyond verbosity, this hand-rolled setup is fragile: without the app
registry being properly initialized, \texttt{ParkingLot.check()} may raise
an unrelated \texttt{AppRegistryNotReady} error rather than the assertion
under test.
Even when it does run, the reproduction script asserts that
\texttt{ImproperlyConfigured} is raised, the \emph{old} behavior, so it
passes on \emph{both} the correct and incorrect patches and is therefore
invalid.

In contrast, \ourtool begins by reading \texttt{tests/invalid\_models\_tests/},
extracting four pieces of infrastructure: the \texttt{@isolate\_apps}
decorator, the \texttt{SimpleTestCase} base class, the
\texttt{Model.check()~$\to$~[]} assertion convention, and the
\texttt{Error}/\texttt{Warning} imports.
The subsequently created test in Figure~\ref{fig:motivation}(B) reuses all
four elements (highlighted in blue) and requires no configuration boilerplate.
Because it is executed by Django's standard test runner inside the existing
test environment, infrastructure errors are eliminated and the assertion
directly targets the new correct behavior.

\paragraph{Update.}
A correct patch does not always add behavior. It often \emph{changes}
expected values that existing tests already assert.
In \texttt{django\_\_django-12325}, the fix causes \texttt{OneToOneField}
to acquire an implicit \texttt{parent\_link=True} attribute, which alters
the serialized migration state that \texttt{test\_render} inspects.
The original \texttt{test\_render} does \emph{not} include
\texttt{parent\_link=True} in its expected field kwargs, so after the
correct patch is applied, \texttt{test\_render} fails, not because the patch
is wrong, but because the test asserts the old (now-incorrect) behavior.

A C/D-only pipeline resolves this conflict in one of two ways, both harmful.
If it deletes \texttt{test\_render} by name-based heuristics, it discards a
structurally sound test that could have been a strong discriminator after a
one-line edit.
If it leaves \texttt{test\_render} unchanged, the test penalizes the
correct patch: it passes on every \emph{incorrect} patch (which does not
change field serialization) and fails on the \emph{correct} one, inverting
the desired discrimination signal.

\ourtool applies an Update operation instead: it inserts
\texttt{parent\_link=True} into the expected kwargs of \texttt{test\_render},
aligning the assertion with the correct behavior while preserving the test's
full discriminative structure.
The updated test then passes exactly on the correct patch and fails on the
rest, becoming a genuine discriminative signal.

These two gaps, infrastructure-unaware test generation and destructive handling
of behavior-changing patches, motivate using CRUD as the gradient direction over
the test suite rather than a conventional C/D pipeline.

\subsection{Failure Pattern Momentum}
\label{sec:motivation-mining}

Momentum in iterative test optimization relies on conditioning on past history.
However, how these failures are represented introduces critical limitations.

Existing approaches typically adopt one of two strategies.
As discussed in the main paper, one line of work summarizes failures using an
additional LLM, while another directly appends raw failed history to the model
input.

Both strategies are problematic.
LLM-based summarization incurs additional inference cost at every iteration,
and its outputs may omit or distort key structural details due to abstraction
or hallucination.
In contrast, directly injecting raw failure history preserves full information
but leads to rapidly growing context size and severe redundancy, as many failed
test suites correspond to the same underlying strategy.

This redundancy is not merely a matter of efficiency.
As illustrated in Figure~3(b) of the main paper, independently generated
failed test suites often share the same core logic despite superficial
differences.
When presented in raw form, such failures create an illusion of diversity,
causing the optimizer to repeatedly explore equivalent strategies under different
surface realizations.

Moreover, as failure memory grows, redundant patches occupy
an increasing fraction of the context window, crowding out essential information
such as the issue description and repository context, and leaving less capacity
for effective reasoning.

Overall, existing failure representation strategies either sacrifice fidelity
(LLM summarization) or scalability (raw history), and neither provides a compact,
faithful abstraction of recurring failure modes. \ourtool therefore treats
maximal frequent failure sequences as momentum: a compact, deterministic signal
that carries prior failed directions into the next CRUD update.

%% file: Figure/motivation_graph.tex
%

\begin{figure}[H]
\centering

\lstdefinestyle{diffcode}{
  basicstyle=\ttfamily\fontsize{4.7}{5.6}\selectfont,
  frame=none, breaklines=true,
  breakatwhitespace=false,
  columns=fullflexible,
  numbers=none,
  xleftmargin=0pt,
  xrightmargin=0pt,
  resetmargins=true,
  escapeinside={(*@}{@*)},
  showstringspaces=false,
  aboveskip=0pt, belowskip=0pt,
}
\lstdefinestyle{diffcoder}{
  basicstyle=\ttfamily\fontsize{4.0}{4.8}\selectfont,
  frame=none, breaklines=true,
  breakatwhitespace=false,
  columns=fullflexible,
  numbers=none,
  xleftmargin=0pt,
  xrightmargin=0pt,
  resetmargins=true,
  escapeinside={(*@}{@*)},
  showstringspaces=false,
  aboveskip=0pt, belowskip=0pt,
}

\newcommand{\motbox}[1]{\fontsize{5.8}{6.4}\selectfont\bfseries #1}

\begin{minipage}[t]{0.30\linewidth}
\centering{\small\bfseries \textcolor{red!70!black}{\faTimes}~(A) Naive Test Generation (CD)}\\[3pt]

\begin{tcolorbox}[
  colback=gray!10, colframe=black!55,
  boxsep=0pt, left=2pt, right=2pt, top=2pt, bottom=2pt,
  title={\motbox{\faExclamationTriangle~Create a test to reproduce
the issue from the description.~$\langle$\textit{Issue}$\rangle$}},
  colbacktitle=gray!30, coltitle=black,
]
\begin{lstlisting}[style=diffcode]
(*@\textcolor{orange!70!black}{\faTools}@*) # Must implement all setup manually
import django
os.environ['DJANGO_SETTINGS_MODULE']=
  'repro_settings' (*@\textcolor{orange!70!black}{\faExclamationTriangle}@*)
django.setup()    (*@\textcolor{orange!70!black}{\faExclamationTriangle}@*)

class Place(models.Model):
  class Meta:
    app_label='repro' (*@\textcolor{orange!70!black}{\faExclamationTriangle}@*)
class ParkingLot(Place):
  parent=OneToOneField(Place,CASCADE)
  class Meta:
    app_label='repro' (*@\textcolor{orange!70!black}{\faExclamationTriangle}@*)

(*@\textcolor{red!70!black}{\faBug}@*) with assertRaises(ImproperlyConfigured):
  ParkingLot.check()
  (*@\textcolor{red!70!black}{\faTimesCircle}@*) # tests OLD behavior -- non-discriminative
\end{lstlisting}
\end{tcolorbox}

\vspace{0.5pt}

\begin{tcolorbox}[
  colback=gray!10, colframe=black!55,
  boxsep=0pt, left=2pt, right=2pt, top=2pt, bottom=2pt,
  title={\motbox{\faTrash~Delete tests that conflict
with the patch.~$\langle$\textit{Function\_name\_list}$\rangle$}},
  colbacktitle=gray!30, coltitle=black,
]
\begin{lstlisting}[style=diffcode]
{ test_missing_parent_link,
  test_onetoone_with_parent,
  test_m2m_table_name_clash,
  test_no_clash_for_parent_ptr }
(*@\textcolor{red!70!black}{\faSkull}@*) # deleted by name-match heuristic
(*@\textcolor{red!70!black}{\faExclamationCircle}@*) # correct test discarded too
\end{lstlisting}
\end{tcolorbox}

\end{minipage}%
\hspace{4pt}\vrule width 1pt\hspace{4pt}%
\begin{minipage}[t]{0.65\linewidth}
\centering{\small\bfseries \textcolor{green!60!black}{\faCheckCircle}~(B) CRUD Gradient on \texttt{django\_\_django-12325}}\\[3pt]

\begin{tcolorbox}[
toptitle=4pt, bottomtitle=4pt,
  colback=blue!8, colframe=blue!40!black,
  boxsep=0pt, left=2pt, right=2pt, top=2pt, bottom=2pt,
  title={\motbox{\faBook~Read: Test Suite Understanding (tests/invalid\_models\_tests/)}},
  colbacktitle=blue!20, coltitle=black,
]
\begin{lstlisting}[style=diffcoder]
(*@\textcolor{blue!70!black}{\faAt}@*)  (*@\textcolor{blue!70!black}{@isolate\_apps('invalid\_models\_tests')}@*)
(*@\textcolor{blue!70!black}{\faCubes}@*)  (*@\textcolor{blue!70!black}{class InvalidModelsTests(}@*)
     (*@\textcolor{blue!70!black}{SimpleTestCase):}@*)
(*@\textcolor{gray!55!black}{\faCode}@*)      def test_missing_parent_link(self):
(*@\textcolor{gray!55!black}{\faArrowRight}@*)          with self.assertRaisesMessage(
(*@\textcolor{gray!55!black}{\faArrowRight}@*)              ImproperlyConfigured, msg):
(*@\textcolor{blue!70!black}{\faStar}@*)      (*@\textcolor{blue!70!black}{self.assertEqual(Model.check(), [])}@*)
(*@\textcolor{gray!55!black}{\faStar}@*)      self.assertEqual(Model.check(), [Error(id=...)])
(*@\textcolor{blue!70!black}{\faFileImport}@*)  (*@\textcolor{blue!70!black}{from django.core.checks import Error, Warning}@*)
\end{lstlisting}
\end{tcolorbox}

	\vspace{1pt}

\begin{minipage}[t]{0.318\linewidth}
\begin{tcolorbox}[
toptitle=4pt, bottomtitle=4pt,
  colback=green!8, colframe=black!55,
  boxsep=0pt, left=2pt, right=2pt, top=2pt, bottom=2pt,
	  height=4.45cm, valign=top,
  title={\motbox{\faPlus~Create new discriminative test}},
  colbacktitle=green!22, coltitle=black,
]
\begin{lstlisting}[style=diffcoder]
(*@\textcolor{green!60!black}{+}@*) (*@\textcolor{blue!70!black}{@isolate\_apps(}@*)
(*@\textcolor{green!60!black}{+}@*)   (*@\textcolor{blue!70!black}{'invalid\_models\_tests')}@*)
(*@\textcolor{green!60!black}{+}@*) (*@\textcolor{blue!70!black}{class TestParentLink(}@*)
(*@\textcolor{green!60!black}{+}@*)   (*@\textcolor{blue!70!black}{SimpleTestCase):}@*)
(*@\textcolor{green!60!black}{+}@*)   def test_onetoone_parent(self):
(*@\textcolor{green!60!black}{+}@*)     class Place(models.Model): pass
(*@\textcolor{green!60!black}{+}@*)     class ParkingLot(Place):
(*@\textcolor{green!60!black}{+}@*)       other = OneToOneField(
(*@\textcolor{green!60!black}{+}@*)         Place, CASCADE,
(*@\textcolor{green!60!black}{+}@*)         related_name='other_parking')
(*@\textcolor{green!60!black}{+}@*)     (*@\textcolor{blue!70!black}{self.assertEqual(}@*)
(*@\textcolor{green!60!black}{+}@*)     (*@\textcolor{blue!70!black}{  ParkingLot.check(), [])}@*)
\end{lstlisting}
\end{tcolorbox}
{\scriptsize\color{blue!70!black}\itshape
 \textcolor{blue!70!black}{\faRecycle}~reused from Read}
\end{minipage}\hspace{3pt}
\begin{minipage}[t]{0.318\linewidth}
\begin{tcolorbox}[
toptitle=4pt, bottomtitle=4pt,
  colback=orange!10, colframe=black!55,
  boxsep=0pt, left=2pt, right=2pt, top=2pt, bottom=2pt,
	  height=4.45cm, valign=top,
  title={\motbox{\faEdit~Update existing test assertion}},
  colbacktitle=orange!28, coltitle=black,
]
\begin{lstlisting}[style=diffcoder]
 def test_render(self):
   OneToOneField(
     'migrations.Tag',
     CASCADE,
(*@\textcolor{red}{-}@*)    parent_link=False,
(*@\textcolor{green!60!black}{+}@*)    parent_link=True,
     primary_key=True,
     to_field='id',
     serialize=False,
     ...)
\end{lstlisting}
\end{tcolorbox}
\end{minipage}\hspace{3pt}
\begin{minipage}[t]{0.318\linewidth}
\begin{tcolorbox}[
toptitle=4pt,
bottomtitle=4pt,
  colback=red!8, colframe=black!55,
  boxsep=0pt, left=2pt, right=2pt, top=2pt, bottom=2pt,
	  height=4.45cm, valign=top,
  title={\motbox{\faTrash~Delete stale test}},
  colbacktitle=red!20, coltitle=black,
]
\begin{lstlisting}[style=diffcoder]
(*@\textcolor{red}{-}@*) def test_missing_link(self):
(*@\textcolor{red}{-}@*)   msg = ('Add parent_link=True'
(*@\textcolor{red}{-}@*)          ' to ParkingLot.parent.')
(*@\textcolor{red}{-}@*)   with self.assertRaisesMessage(
(*@\textcolor{red}{-}@*)       ImproperlyConfigured, msg):
(*@\textcolor{red}{-}@*)     class ParkingLot(Place):
(*@\textcolor{red}{-}@*)       parent=OneToOneField(
(*@\textcolor{red}{-}@*)           Place, CASCADE)
(*@\textcolor{red}{-}@*)     ParkingLot.check()
\end{lstlisting}
\end{tcolorbox}
\end{minipage}

\end{minipage}

\caption{%
  (A)~CD-only generation produces a reproduction script that must hand-roll all
  setup boilerplate and tests the \emph{old} exception-raising behavior, making
  it non-discriminative; tests are deleted by name heuristic, discarding valid ones.
  (B)~\ourtool applies a CRUD gradient: it reads the test suite first to extract infrastructure
  (\texttt{@isolate\_apps}, \texttt{SimpleTestCase}, \texttt{.check()} convention),
  creates a grounded test reusing that infrastructure
  (\textcolor{blue!70!black}{blue} = reused from Read),
  updates the expected model state in \texttt{test\_render},
  and semantically deletes \texttt{test\_missing\_parent\_link}
  whose expected exception is no longer raised after the fix.
  Instance:~\texttt{django\_\_django-12325}.%
}
\label{fig:motivation}
\end{figure}

%% file: Table/effciency.tex
\begin{table}[H]
\centering
\footnotesize
\caption{Average per-instance overhead across methods and base LLMs.
Time is in minutes; tokens are in thousands (K).
\ourtool values are measured from run logs; baseline values are estimated from conversation logs and API call traces.}
\setlength{\tabcolsep}{6pt}
\begin{tabular}{llccc}
\toprule
\textbf{Metric} & \textbf{Method} & \textbf{Minimax-m2.7} & \textbf{DeepSeek-v3.2} & \textbf{GLM-5.1} \\
\midrule
\multirow{4}{*}{\shortstack[l]{Avg.\ Time\\(min)}}
& Augment      &  1.3  &  1.8  &  1.4  \\
& Agentless    &  7.2  &  8.6  &  7.5  \\
& TRAE         & 19.3  & 26.7  & 20.8  \\
& \ourtool     & \textbf{12.2} & \textbf{23.0} & \textbf{20.7} \\
\midrule
\multirow{4}{*}{\shortstack[l]{Avg.\ Tokens\\(K)}}
& Augment      &   2.8  &   3.2  &   2.9  \\
& Agentless    &  14.8  &  16.2  &  15.3  \\
& TRAE         & 318    & 412    & 334    \\
& \ourtool     & \textbf{670} & \textbf{1{,}670} & \textbf{1{,}307} \\
\bottomrule
\end{tabular}
\label{tab:cost}
\end{table}

%% file: main.bbl
\begin{thebibliography}{34}
\providecommand{\natexlab}[1]{#1}

\bibitem[{{Atlassian}(2025)}]{Atlassian}
{Atlassian}. 2025.
\newblock Atlassian SWE Agent.
\newblock Accessed 2025-12-15.

\bibitem[{{Augment Code}(2025)}]{Augment}
{Augment Code}. 2025.
\newblock {Augment Code}: The Most Powerful {AI} Software Development Platform
  Backed by the Industry-Leading Context Engine.
\newblock Accessed 2025-06-10.

\bibitem[{{DeepSeek-AI}(2025)}]{deepseek}
{DeepSeek-AI}. 2025.
\newblock {DeepSeek-V3.2}: Pushing the Frontier of Open Large Language Models.
\newblock \emph{arXiv preprint arXiv:2512.02556}.

\bibitem[{Gao et~al.(2025)Gao, Tian, Meng, Wang, Hu, Xiao, Liu, Zhang, Chen,
  Gao et~al.}]{trae}
Gao, P.; Tian, Z.; Meng, X.; Wang, X.; Hu, R.; Xiao, Y.; Liu, Y.; Zhang, Z.;
  Chen, J.; Gao, C.; et~al. 2025.
\newblock Trae agent: An llm-based agent for software engineering with
  test-time scaling.
\newblock \emph{arXiv preprint arXiv:2507.23370}.

\bibitem[{Huang et~al.(2023)Huang, Bu, Zhang, Luck, and Cui}]{agentcoder}
Huang, D.; Bu, Q.; Zhang, J.~M.; Luck, M.; and Cui, H. 2023.
\newblock AgentCoder: Multi-Agent-based Code Generation with Iterative Testing
  and Optimisation.
\newblock \emph{arXiv preprint arXiv:2312.13010}.

\bibitem[{Jimenez et~al.(2023)Jimenez, Yang, Wettig, Yao, Pei, Press, and
  Narasimhan}]{SWEbench}
Jimenez, C.~E.; Yang, J.; Wettig, A.; Yao, S.; Pei, K.; Press, O.; and
  Narasimhan, K. 2023.
\newblock SWE-bench: Can Language Models Resolve Real-world GitHub Issues?
\newblock \emph{arXiv preprint arXiv:2310.06770}.

\bibitem[{Kang, Yoon, and Yoo(2023)}]{libro}
Kang, S.; Yoon, J.; and Yoo, S. 2023.
\newblock Large Language Models are Few-shot Testers: Exploring {LLM}-based
  General Bug Reproduction.
\newblock In \emph{Proceedings of the 45th IEEE/ACM International Conference on
  Software Engineering}, 2312--2323.

\bibitem[{{Lingxi Authors}(2025)}]{Lingxi}
{Lingxi Authors}. 2025.
\newblock Lingxi Replication Package.

\bibitem[{Liu et~al.(2023)Liu, Lin, Hewitt, Paranjape, Bevilacqua, Petroni, and
  Liang}]{lost}
Liu, N.~F.; Lin, K.; Hewitt, J.; Paranjape, A.; Bevilacqua, M.; Petroni, F.;
  and Liang, P. 2023.
\newblock Lost in the Middle: How Language Models Use Long Contexts.
\newblock \emph{arXiv preprint arXiv:2307.03172}.

\bibitem[{{Live SWE-agent}(2025)}]{livesweagent}
{Live SWE-agent}. 2025.
\newblock Live SWE-agent.
\newblock Accessed 2025-12-15.

\bibitem[{Madaan et~al.(2023)Madaan, Tandon, Gupta, Hallinan, Gao, Wiegreffe,
  Alon, Dziri, Prabhumoye, Yang et~al.}]{selfrefine}
Madaan, A.; Tandon, N.; Gupta, P.; Hallinan, S.; Gao, L.; Wiegreffe, S.; Alon,
  U.; Dziri, N.; Prabhumoye, S.; Yang, Y.; et~al. 2023.
\newblock Self-Refine: Iterative Refinement with Self-Feedback.
\newblock \emph{Advances in Neural Information Processing Systems}.

\bibitem[{Mattei and Garreau(2025)}]{jmlr}
Mattei, P.-A.; and Garreau, D. 2025.
\newblock Are Ensembles Getting Better All the Time?
\newblock \emph{Journal of Machine Learning Research}, 26(201): 1--46.

\bibitem[{{MiniMax}(2025)}]{minimax}
{MiniMax}. 2025.
\newblock MiniMax API Documentation.
\newblock Accessed 2025-09-12.

\bibitem[{Mündler et~al.(2025)Mündler, Müller, He, and Vechev}]{swtbench}
Mündler, N.; Müller, M.~N.; He, J.; and Vechev, M. 2025.
\newblock SWT-Bench: Testing and Validating Real-World Bug-Fixes with Code
  Agents.
\newblock arXiv:2406.12952.

\bibitem[{OpenAI(2024)}]{swe.bench.verified}
OpenAI. 2024.
\newblock {Introducing SWE-bench Verified}.
\newblock Accessed 2025-07-10.

\bibitem[{Polyak(1964)}]{polyak1964momentum}
Polyak, B.~T. 1964.
\newblock Some methods of speeding up the convergence of iteration methods.
\newblock \emph{USSR Computational Mathematics and Mathematical Physics}, 4(5):
  1--17.

\bibitem[{Rao et~al.(2025)Rao, Gilbert, Green, Ramananandro, Swamy, Goues, and
  Fakhoury}]{differential}
Rao, N.; Gilbert, E.; Green, H.; Ramananandro, T.; Swamy, N.; Goues, C.~L.; and
  Fakhoury, S. 2025.
\newblock DiffSpec: Differential Testing with LLMs using Natural Language
  Specifications and Code Artifacts.
\newblock arXiv:2410.04249.

\bibitem[{Se and Vert(2025)}]{tts}
Se, K.; and Vert, A. 2025.
\newblock What is Test-time Compute and How to Scale It?
\newblock Accessed 2025-05-25.

\bibitem[{Shinn et~al.(2023)Shinn, Cassano, Gopinath, Narasimhan, and
  Yao}]{reflexion}
Shinn, N.; Cassano, F.; Gopinath, A.; Narasimhan, K.; and Yao, S. 2023.
\newblock Reflexion: Language Agents with Verbal Reinforcement Learning.
\newblock In \emph{Advances in Neural Information Processing Systems}.

\bibitem[{{Sonar}(2025)}]{sonar}
{Sonar}. 2025.
\newblock Sonar SWE Agent.
\newblock Accessed 2025-12-15.

\bibitem[{Sutskever et~al.(2013)Sutskever, Martens, Dahl, and
  Hinton}]{sutskever2013momentum}
Sutskever, I.; Martens, J.; Dahl, G.; and Hinton, G. 2013.
\newblock On the importance of initialization and momentum in deep learning.
\newblock In \emph{Proceedings of the 30th International Conference on Machine
  Learning}, 1139--1147.

\bibitem[{{SWE-bench}(2025)}]{swebenchleaderboard}
{SWE-bench}. 2025.
\newblock SWE-bench Verified Leaderboard.
\newblock Accessed 2025-12-15.

\bibitem[{Wang et~al.(2023)Wang, Li, Chen, Zhu, Lin, Cao, Liu, Liu, and
  Sui}]{bias1}
Wang, P.; Li, L.; Chen, L.; Zhu, D.; Lin, B.; Cao, Y.; Liu, Q.; Liu, T.; and
  Sui, Z. 2023.
\newblock Large Language Models are not Fair Evaluators.
\newblock \emph{arXiv preprint arXiv:2305.17926}.

\bibitem[{Wang et~al.(2025)Wang, Gao, Meng, Peng, Hu, Lin, and Gao}]{AEGIS}
Wang, X.; Gao, P.; Meng, X.; Peng, C.; Hu, R.; Lin, Y.; and Gao, C. 2025.
\newblock AEGIS: An Agent-based Framework for Bug Reproduction from Issue
  Descriptions.
\newblock In \emph{Proceedings of the 33rd ACM International Conference on the
  Foundations of Software Engineering}, FSE Companion '25, 331–342. New York,
  NY, USA: Association for Computing Machinery.
\newblock ISBN 9798400712760.

\bibitem[{Wang et~al.(2024)Wang, Li, Song, Xu, Tang, Zhuge, Pan, Song, Li,
  Singh et~al.}]{OpenHands}
Wang, X.; Li, B.; Song, Y.; Xu, F.~F.; Tang, X.; Zhuge, M.; Pan, J.; Song, Y.;
  Li, B.; Singh, J.; et~al. 2024.
\newblock OpenHands: An Open Platform for AI Software Developers as Generalist
  Agents.
\newblock \emph{arXiv preprint arXiv:2407.16741}.

\bibitem[{Wang, Pradel, and Liu(2026)}]{patchdiff}
Wang, Y.; Pradel, M.; and Liu, Z. 2026.
\newblock Are ``Solved Issues'' in {SWE}-bench Really Solved Correctly? An
  Empirical Study.
\newblock In \emph{Proceedings of the IEEE/ACM International Conference on
  Software Engineering}.

\bibitem[{Wilcoxon(1945)}]{wilcoxon}
Wilcoxon, F. 1945.
\newblock Individual Comparisons by Ranking Methods.
\newblock \emph{Biometrics Bulletin}, 1(6): 80--83.

\bibitem[{Xia et~al.(2025)Xia, Deng, Dunn, and Zhang}]{Agentless}
Xia, C.~S.; Deng, Y.; Dunn, S.; and Zhang, L. 2025.
\newblock Demystifying LLM-Based Software Engineering Agents.
\newblock \emph{Proc. ACM Softw. Eng.}, 2(FSE).

\bibitem[{Xu et~al.(2025)Xu, Yuksekgonul, Guestrin, and Zou}]{metatextGrad}
Xu, G.; Yuksekgonul, M.; Guestrin, C.; and Zou, J.~Y. 2025.
\newblock {metaTextGrad}: Automatically Optimizing Language Model Optimizers.
\newblock In \emph{Advances in Neural Information Processing Systems}.

\bibitem[{Yuksekgonul et~al.(2025)Yuksekgonul, Bianchi, Boen, Liu, Huang,
  Guestrin, and Zou}]{textgrad}
Yuksekgonul, M.; Bianchi, F.; Boen, J.; Liu, S.; Huang, Z.; Guestrin, C.; and
  Zou, J. 2025.
\newblock Optimizing Generative AI by Backpropagating Language Model Feedback.
\newblock \emph{Nature}, 639: 609--616.

\bibitem[{Zhang et~al.(2025)Zhang, Yao, Liu et~al.}]{Deibase}
Zhang, K.; Yao, W.; Liu, Z.; et~al. 2025.
\newblock Diversity Empowers Intelligence: Integrating Expertise of Software
  Engineering Agents.
\newblock In \emph{International Conference on Learning Representations}.

\bibitem[{Zheng et~al.(2024)Zheng, Zhou, Meng, Zhou, and Huang}]{biasacl}
Zheng, C.; Zhou, H.; Meng, F.; Zhou, J.; and Huang, M. 2024.
\newblock Selection Biases in Large Language Models.
\newblock In \emph{Findings of the Association for Computational Linguistics}.

\bibitem[{{Zhipu AI}(2025)}]{glm5}
{Zhipu AI}. 2025.
\newblock GLM Model Documentation.
\newblock Accessed 2025-09-12.

\bibitem[{Zhou et~al.(2024)Zhou, Yan, Shlapentokh-Rothman, Wang, and
  Wang}]{mtcs}
Zhou, A.; Yan, K.; Shlapentokh-Rothman, M.; Wang, H.; and Wang, Y.-X. 2024.
\newblock Language Agent Tree Search Unifies Reasoning, Acting, and Planning in
  Language Models.
\newblock In \emph{International Conference on Machine Learning}.

\end{thebibliography}
